\documentstyle[revtex]{aps}
\newtheorem{theorem}{Theorem}
\begin{document}
\draft
\begin{title}
Self-consistent anisotropic oscillator with cranked angular and vortex
velocities
\end{title}
\author{G.\ Rosensteel}
\begin{instit}
Department of Physics, Tulane University, New Orleans, Louisiana 70118
\end{instit}
\begin{abstract}
Rotating deformed nuclei are neither rigid rotors nor irrotational droplets.
The
Kelvin circulation vector is the kinematical observable that measures the true
character of nuclear rotation.  For the anisotropic oscillator
potential, mean field
solutions with fixed angular momentum L and Kelvin circulation C are derived in
analytic form.  The cranking Lagrange multipliers corresponding to the L and C
constraints are the angular $\omega$ and vortex $\lambda$ velocities.
\mbox{Self-consistent} solutions are reported with a constraint to constant
volume.
\end{abstract}
\narrowtext
\section{INTRODUCTION}

The familiar adiabatic rotor model enables the determination of {\em static}
nuclear shapes from collective multipole transition data.  By applying the
Alaga
rules of this simple geometrical model, experimental $E2$ transition rates are
interpreted in terms of the $\beta$ and $\gamma$ shape deformation
parameters.  The Riemann rotor model is a similar enabling model for {\em
dynamical} nuclear currents.\cite{ROS88}  Transverse form factors and other
experimental measures of nuclear collective dynamics may be interpreted in
terms of the rigidity parameter $r$.  This parameter ranges continuously
between the limiting cases of rigid rotation $r=1$ and irrotational flow $r=0$.
The observable measuring the static deformation is the quadrupole operator
$Q^{(2)}$; the vector observables measuring the dynamical current are the
angular momentum $\vec{L}$ and the Kelvin circulation $\vec{C}$.

The first aim of this paper is to show that the classical
expressions\cite{CHA69,LEB67} for the kinetic energy, angular momentum, and
Kelvin circulation of a Riemann rotor may be derived by simultaneous angular
and vortex ``Inglis'' cranking of the quantum anisotropic
oscillator\cite{ING54,RIN80}
\FL
\begin{equation}
H_{\omega \lambda}  = - \frac{\hbar^{2}}{2m} \bigtriangleup +
\frac{1}{2}m(\omega^{2}_{x} x^{2}+\omega^{2}_{y} y^{2}+\omega^{2}_{z} z^{2}) -
(\omega \hat{L}_{x}-\lambda \hat{C}_{x}).
\end{equation}
This semiclassical correspondence is achieved in perturbation theory for small
angular and vortex velocities provided the field is \mbox{self-consistent} with
the
shape,
\begin{equation}
\omega_{x}N_{x} = \omega_{y}N_{y} = \omega_{z}N_{z}, \label{SELF}
\end{equation}
where $N_{k} = \sum (n_{k}+1/2)$ denotes the total number of quanta in the
k$^{th}$ direction.  When the vortex velocity vanishes $\lambda =0$,
 \mbox{self-consistency} implies rigid rotation, a \mbox{well-known} result for
ordinary Inglis cranking of the angular velocity.\cite{BOH75,LUD60}  When
$\lambda \neq 0$, \mbox{self-consistency} implies the Riemann rotor moment of
inertia which is an interpolation between the rigid and irrotational flow
moments.
Next, analytic formulas are discovered for the exact eigenvalues of the
Routhian
$H_{\omega \lambda}$.  Calculations at finite $\omega$ and $\lambda$ are
reported for $^{\rm 20}$Ne and $^{166}$Er with a constraint to constant volume.

To define and interpret physically the vortex velocity and the Kelvin
circulation,
consider a system of $A$ particles and its associated inertia ellipsoid.  With
respect to an inertial \mbox{center-of-mass} frame, the inertia ellipsoid
rotates
with angular velocity $\vec{\omega}$.  By definition, the inertia tensor in the
rotating intrinsic frame is diagonal,
\begin{equation}
Q_{ij} = \sum_{\alpha =1}^{A}x_{\alpha i}x_{\alpha j} =
\frac{A}{5}a_{i}^{2}\delta_{ij},
\end{equation}
where $\vec{x}_{\alpha}$ denotes the coordinates of particle $\alpha$ in the
\mbox{body-fixed} frame and the $a_{i}$ are the axes lengths of the inertia
ellipsoid.  In quantum mechanics, the angular momentum and circulation are
defined by the operators\cite{ROS88}
\begin{eqnarray}
\hat{L}_{k} & = & -i\hbar \sum_{\alpha}\left( x_{\alpha
i}\frac{\partial}{\partial
x_{\alpha j}}-x_{\alpha j}\frac{\partial}{\partial x_{\alpha i}}\right)
\nonumber \\
\hat{C}_{k} & = & -i\hbar \sum_{\alpha}\left( \frac{a_{j}}{a_{i}}x_{\alpha
i}\frac{\partial}{\partial x_{\alpha j}}-\frac{a_{i}}{a_{j}}x_{\alpha
j}\frac{\partial}{\partial x_{\alpha i}}\right) \label{KELOP}\\
         & = & -i\hbar \sum_{\alpha}\left( y_{\alpha
i}\frac{\partial}{\partial y_{\alpha
j}}-y_{\alpha j}\frac{\partial}{\partial y_{\alpha i}}\right) \nonumber ,
\end{eqnarray}
where $y_{\alpha i} = x_{\alpha i}/a_{i}$ are the dimensionless coordinates of
particle $\alpha$ in the stretched intrinsic coordinate system and $i,j,k$ are
cyclic.  In the stretched system, the inertia ellipsoid is transformed
into a sphere
of unit radius, and the inertia tensor is a multiple of the identity matrix.
The
Kelvin circulation is evidently the generator of rotations in this
stretched frame.
Adopting the active viewpoint, the operator $\exp (-i\vec{\omega}\cdot
\vec{L}/\hbar)$ generates a finite rotation of the nuclear system with an
angular
velocity $\vec{\omega}$ with respect to the laboratory frame.  The operator
$\exp
(i\vec{\lambda}\cdot \vec{C}/\hbar)$ generates a vortex rotation of the
nucleons
with respect to the \mbox{body-fixed} frame with the vortex velocity
$\vec{\lambda}$.

In this introductory section, the theory of classical Riemann rotors is
reviewed
briefly.  A classical Riemann rotor is a constant density fluid with an
ellipsoidal
boundary and a velocity field that is a linear function of the position
coordinates.
Classical Riemann rotors provide models for rotating stars and
galaxies.\cite{CHA69,LEB67}  For nuclei, a linear velocity field was proposed
by
Cusson.\cite{CUS68}  For these classical rotors, the uniform vorticity
$\vec{\zeta}$, defined as the curl of the velocity field with respect to the
\mbox{body-fixed} frame, is given by
\begin{equation}
\zeta_{k} = -\frac{a_{i}^{2}+a_{j}^{2}}{a_{i}a_{j}}\lambda_{k}.
\end{equation}
The curl of the laboratory frame velocity field $\vec{U}$ resolved along the
intrinsic axes is the inertial frame vorticity,
\begin{equation}
\vec{\zeta}^{(0)} = \vec{\nabla}_{x}\times \vec{U} = \vec{\zeta} +
2\vec{\omega}.
\end{equation}
Then the Kelvin circulation of a Riemann rotor, defined by the line integral of
the
velocity field around the ellipse $C_{k}$ bounding the fluid in the $i-j$
principal
plane, is given via Stoke's theorem by the expression
\begin{equation}
C_{k} = \frac{M}{5\pi}\oint_{C_{k}}\vec{U}\cdot d\vec{l} =
\frac{M}{5}a_{i}a_{j}(\zeta_{k}+2\omega_{k}),
\end{equation}
since $\pi a_{i}a_{j}$ is the area of the ellipse $C_{k}$.  $M$ is the
fluid's mass.
Note that irrotational flow is attained if the circulation $\vec{C}$
and the inertial
frame vorticity $\vec{\zeta}^{(0)}$ vanish, and the \mbox{body-fixed} uniform
vorticity satisfies $\vec{\zeta}=-2\vec{\omega}$.
For a classical Riemann rotor, the angular momentum and circulation are
\begin{eqnarray}
L_{k} & = & (M/5)\left[ (a_{i}^{2}+a_{j}^{2})\omega_{k} -
2a_{i}a_{j}\lambda_{k}\right] \nonumber \\
C_{k} & = & (M/5)\left[ 2a_{i}a_{j}\omega_{k} -
(a_{i}^{2}+a_{j}^{2})\lambda_{k}\right] \label{expLK}.
\end{eqnarray}
Ignoring vibrations of the axes lengths, the kinetic energy of a classical
Riemann
rotor is a combination of centrifugal and Coriolis terms
\begin{equation}
T(\vec{\omega},\vec{\lambda}) = \frac{1}{2}\left( \vec{\omega}\cdot \vec{L} -
\vec{\lambda}\cdot \vec{C}\right) .
\end{equation}
The angular momentum and circulation are given by derivatives of the kinetic
energy with respect to the angular velocity and the vortex velocity
\begin{equation}
L_{k} = \left( \frac{\partial T}{\partial \omega_{k}}\right)_{\lambda}
\hspace{.2in}
C_{k} = -\left( \frac{\partial T}{\partial
\lambda_{k}}\right)_{\omega}\label{LKDT}.
\end{equation}

A Riemann ellipsoid in equilibrium has constant axes lengths, angular
momentum and Kelvin circulation.  The physical interpretation of equilibrium
Riemann ellipsoids is most transparent for an \mbox{S-type} Riemann rotor that
rotates about one principal axis, say the $x$-axis.  In this case, the angular
momentum, Kelvin circulation, angular velocity, and vortex velocity vectors are
aligned with the $x$-axis and the kinetic energy simplifies to
\begin{equation}
T(\omega ,\lambda ) = \frac{I_{0}}{4}\left\{ (b^2+c^2)(\omega^{2}+\lambda^{2})
-
4bc\omega \lambda \right\} \label{RRKE},
\end{equation}
where $b=a_{2}/R$ and $c=a_{3}/R$ are the $y$ and $z$ axes lengths in units
of a characteristic length R, and the moment of inertia of a sphere of radius R
is
$I_{0}=(2/5)MR^{2}$.

The S-type equilibrium solutions are classified by a single parameter, the {\em
rigidity} $r=1+\zeta/(2\omega )$.  When $r=1$ the vortex velocity is zero and
the
rotation is rigid.  When $r=0$ the circulation vanishes and the velocity field
is
irrotational, because the circulation is directly proportional to the
rigidity and the
cross-sectional area of the bounding ellipse,
\begin{equation}
C = I_{0}\,bc\,r\,\omega \label{Krw}.
\end{equation}
Riemann rotors with $0<r<1$ span the full range of dynamical potentialities
from
irrotational flow to rigid rotation, consistent with the constraint to
a S-type linear
velocity field.  The angular momentum, kinetic energy, and velocity field share
a
remarkable property for S-type ellipsoids:  Each of these quantities is a
simple
convex combination of their corresponding rigid rotor (RR) and irrotational
flow
(IF) values,
\begin{mathletters}
\begin{eqnarray}
L & = & I_{r}\omega \\
T & = & \frac{1}{2}I_{r^{2}}\omega^{2} \\
U & = & r U_{\rm RR} + (1-r) U_{\rm IF},
\end{eqnarray}
\end{mathletters}
where the interpolated inertia is $I_{p}\equiv (p I_{\rm RR}+(1-p) I_{\rm IF})$
for
$0\leq p\leq 1$ and the rigid body and irrotational flow moments of inertia are
given by
\begin{equation}
I_{\rm RR} = \frac{I_{0}}{2}(b^{2}+c^{2}), \mbox{ } I_{\rm
IF}=\frac{I_{0}}{2}\frac{(c^{2}-b^{2})^{2}}{(c^{2}+b^{2})}.
\end{equation}
In terms of the angular momentum, the kinetic energy is
\begin{equation}
T = \frac{L^{2}}{2{\cal I}_{r}}, \mbox{ } {\cal
I}_{r}=\frac{(I_{r})^2}{I_{r^{2}}}.
\label{TL2}
\end{equation}

In the classical theory of fluid dynamics, the net Kelvin circulation
is proved to be
conserved by the hydrodynamic equations of motion.\cite{CHA69,LEB67}
Hence, to the extent that noncollective degrees of freedom may be ignored,
geometrical states forming rotational bands in nuclei are conjectured to be
approximate eigenstates of the Kelvin circulation.

A direct experimental method for determining the nuclear circulation and
rigidity
is provided by inelastic electron scattering measurements of the transverse
$E2$
form factor, as has been emphasized by Moya de Guerra\cite{GUE86} and
Vassanji and Rowe.\cite{VAS87}  The author has shown elsewhere that the
transverse $E2$ form factor for a Riemann rotor is a weighted interpolation of
the rigid rotor and irrotational flow form factors
\FL
\begin{equation}
{\cal F}^{E2}(q) = \left[ r I_{\rm RR}{\cal F}^{E2}_{\rm RR}(q) +
(1-r)I_{\rm IF}{\cal
F}^{E2}_{\rm IF}(q) \right] /I_{r},
\end{equation}
where $\hbar q$ is the momentum transferred in the inelastic electron
scattering.\cite{ROS90a}

There is to date no published experimental measurement of transverse form
factors in the heavy  deformed  region.  However, a method based on
measurement of multiple real photons in coincidence with  the  scattered
electron was proposed recently that may  permit  the  separation  of  the
transverse from the longitudinal component due to interference terms  in
the angular distribution formulae \cite{GUE90}.  If data were available, one
would fit the expression for the Riemann form factor to the first peak and,
thereby, measure the experimental rigidity $r$.

The projected Hartree-Fock calculations of the transverse form factor in the
heavy deformed region made by Berdichevsky {\it et al.\ }\cite{GUE88}
enable a theoretical estimate for the rigidity.  For $^{156}$Gd, these PHF
calculations correspond to a rigidity $r \sim 0.12$.  This value compares
favorably with the value predicted via Eq.\  (\ref{TL2}) from the $^{156}$Gd
measured moment of inertia and deformation, $r \sim 0.15$.

This application of the Riemann model to dynamical currents is similar to the
application of the adiabatic rotor model to static shapes.  The physical
meaning
of a B(E2) transition rate, either measured experimentally or calculated in a
detailed theoretical model, is obtained by its rotor model interpretation in
terms
of the nuclear shape, i.e., $\beta$ and $\gamma$ parameters.  Similarly, the
Riemann model attaches physical meaning to measurements and calculations of
transverse $E2$ form factors in terms of the Kelvin circulation and the
rigidity
parameter $r$.

\section{CRANKED ANISOTROPIC OSCILLATOR}

\subsection{Semiclassical Correspondence}
To solve the energy eigenvalue problem for the cranked anisotropic
oscillator
Hamiltonian, introduce the oscillator creation and annihilation bosons and
\mbox{re-express} the single-particle Kelvin circulation of Eq.\  (\ref{KELOP})
as
\begin{eqnarray}
\hat{C}_{x} & = & \frac{-i\hbar}{2\sqrt{\omega_{y}\omega_{z}}} \left\{
(\frac{c}{b}\omega_{z}+\frac{b}{c}\omega_{y})(c_{y}^{\dagger}c_{z}-
c_{z}^{\dagger}c_{y}) \right. \nonumber \\
& & \mbox{    } \left.  + (\frac{b}{c}\omega_{y}-
\frac{c}{b}\omega_{z})(c_{y}^{\dagger}c_{z}^{\dagger}-c_{z}c_{y})\right\} .
\end{eqnarray}
The circulation operator reduces to the angular momentum operator
$\hat{L}_{x}$, if the axes lengths are replaced by unity.  Substituting these
expressions for the Kelvin circulation and angular momentum operators, the
single-particle Routhian for the anisotropic oscillator potential is
written in terms
of bosons as
\FL
\begin{eqnarray}
H_{\omega \lambda} & = & \hbar \omega_{x}(c_{x}^{\dagger}c_{x}+\frac{1}{2})
+ \hbar \omega_{y}(c_{y}^{\dagger}c_{y}+\frac{1}{2}) + \hbar
\omega_{z}(c_{z}^{\dagger}c_{z}+\frac{1}{2}) \nonumber \\
& & \mbox{} - (\omega \hat{L}_{x}-\lambda \hat{C}_{x}).
\end{eqnarray}
In perturbation theory, for small angular and vortex velocities, the collective
kinetic energy T($\omega$,$\lambda$) of the $A$-nucleon system is given by
Inglis's cranking formula:\cite{ING54}
\FL
\begin{eqnarray}
T & = & \sum_{ph}\frac{\mid \langle p \mid \omega \hat{L}_{x}-
\lambda \hat{C}_{x}\mid h \rangle \mid^{2}}{\epsilon_{p}-\epsilon_{h}}
\nonumber
\\
& = & \frac{\hbar}{4\omega_{y}\omega_{z}}\left\{\left| \omega
(\omega_{y}+\omega_{z})-\lambda (\frac{b}{c}\omega_{y}+\frac{c}{b}\omega_{z})
\right|^{2} \frac{N_{z}-N_{y}}{\omega_{y}-\omega_{z}} \right. \nonumber \\
& & \mbox{} \left. + \left| \omega (\omega_{y}-\omega_{z})-\lambda
(\frac{b}{c}\omega_{y}-\frac{c}{b}\omega_{z}) \right|^{2}
\frac{N_{z}+N_{y}}{\omega_{y}+\omega_{z}} \right\} . \label{INGLIS}
\end{eqnarray}
The second half of this equation is proven by applying the same techniques that
work for the $\lambda =0$ case.\cite{RIN80}  Also, in perturbation theory, the
expectations of the axes lengths are given by
\begin{equation}
\frac{a^{2}}{5}R^{2} = \frac{1}{A}\langle \sum_{\alpha =1}^{A}x_{\alpha}^{2}
\rangle = \frac{\hbar}{m A}\frac{N_{x}}{\omega_{x}}, \label{AXESexp}
\end{equation}
and similarly for $b^{2}$ and $c^{2}$.  Self-consistency of the shape of the
potential field with the spatial density distribution requires equality
for the ratios
\begin{equation}
a : b : c = \frac{1}{\omega_{x}} : \frac{1}{\omega_{y}} : \frac{1}{\omega_{z}}
,
\end{equation}
viz., Eq.\  (\ref{SELF}).  A principal result of this paper is the following
semiclassical correspondence for the cranked anisotropic oscillator:

\begin{theorem}  For self-consistent perturbation solutions to the cranked
anisotropic oscillator $H_{\omega \lambda}$, the Inglis cranking energy, Eq.\
(\ref{INGLIS}), equals the classical Riemann rotor energy, Eq.\  (\ref{RRKE}).
The expectations of the angular momentum and the Kelvin circulation are given
by their classical values too, Eq.\ (\ref{expLK}), and satisfy the derivative
conditions, Eq.\ (\ref{LKDT}).
\end{theorem}

This theorem is proved by using the self-consistency relation and the formulae
for the expectations of the axes lengths, Eq.\  (\ref{AXESexp}), to eliminate
the
total number of quanta $N_{i}$ and the frequencies $\omega_{i}$  from the
perturbation expressions for the energy eigenvalue and for the expectations of
the angular momentum and Kelvin circulation operators.

\subsection{Analytic Quantum Mean-Field Results}

The Routhian eigenvalue problem may be solved analytically by making a
canonical transformation from the original oscillator bosons to new bosons that
diagonalize $H_{\omega \lambda}$.  This transformation exists because the
Routhian is a quadratic form in the oscillator bosons.\cite{RIP86,RIP75}  The
exact solution to the rigid rotor $\lambda =0$ eigenvalue problem is known
already.\cite{STA78}  For $\lambda \neq 0$, the eigenvalues of the A-particle
Routhian are given by
\begin{equation}
\tilde{E} = \hbar \omega_{x}N_{x} + \hbar \Omega_{+}N_{y} + \hbar \Omega_{-
}N_{z},
\end{equation}
where the frequencies are
\begin{equation}
\Omega_{\pm}^{2} = \frac{1}{2}(\omega_{y}^{2}+\omega_{z}^{2}) + \omega^{2} +
\lambda^{2} - (\frac{c}{b}+\frac{b}{c})\omega \lambda \pm q
\end{equation}
and
\begin{eqnarray}
q^{2} & = & \frac{1}{4}(\omega_{y}^{2}-\omega_{z}^{2})^{2} +
(\omega_{y}^{2}+\omega_{z}^{2})(2\omega^{2}+\lambda^{2}) \nonumber \\
& & \mbox{} + (\frac{b^{2}}{c^{2}}\omega_{y}^{2}+\frac{c^{2}}{b^{2}}
\omega_{z}^{2})\lambda^{2} \nonumber \\
& & \mbox{} - ((\frac{c}{b}+3\frac{b}{c})\omega_{y}^{2} +
(\frac{b}{c}+3\frac{c}{b})\omega_{z}^{2})\omega \lambda .
\end{eqnarray}
Note that in the limit of null cranking,
\begin{equation}
\lim_{\omega ,\lambda \rightarrow 0} \Omega_{\pm} = \omega_{\stackrel{y}{z}}.
\end{equation}

The expectations of the position operators, angular momentum, and circulation
may be calculated by using Feynman's lemma that identifies the expectation of a
derivative of the Hamiltonian operator to the corresponding derivative of the
energy expectation.\cite{FEY39}  For example, the expectation $\langle x^{2}
\rangle$ is determined by
\begin{eqnarray}
\langle \frac{\partial H_{\omega \lambda}}{\partial \omega_{x}} \rangle & = &
\frac{\partial E}{\partial \omega_{x}} \nonumber \\
m \omega_{x}\langle \sum_{\alpha =1}^{A}x_{\alpha}^{2} \rangle & = & \hbar
N_{x} \nonumber \\
\frac{I_{0}}{2 \hbar} a^{2} & = & \frac{N_{x}}{\omega_{x}}.
\end{eqnarray}
The expectations of the other position operators are determined {\em mutatis
mutandis}:
\widetext
\begin{eqnarray}
\frac{I_{0}}{2 \hbar} b^{2} & = &
\frac{1}{2}(\frac{N_{y}}{\Omega_{+}}+\frac{N_{z}}{\Omega_{-}}) +
\frac{1}{2q}(\frac{N_{y}}{\Omega_{+}}-\frac{N_{z}}{\Omega_{-}})
\left[ \frac{1}{2}(\omega_{y}^{2}-\omega_{z}^{2}) +
2\omega^{2} + (1+\frac{b^{2}}{c^{2}})\lambda^{2} -
(\frac{c}{b}+3\frac{b}{c})\omega \lambda \right] \nonumber \\
\frac{I_{0}}{2 \hbar} c^{2} & = &
\frac{1}{2}(\frac{N_{y}}{\Omega_{+}}+\frac{N_{z}}{\Omega_{-}}) -
\frac{1}{2q}(\frac{N_{y}}{\Omega_{+}}-\frac{N_{z}}{\Omega_{-}})
\left[ \frac{1}{2}(\omega_{y}^{2}-\omega_{z}^{2}) -
2\omega^{2} - (1+\frac{c^{2}}{b^{2}})\lambda^{2} +
(\frac{b}{c}+3\frac{c}{b})\omega \lambda \right] . \label{AXES}
\end{eqnarray}
The expectations of the angular momentum and circulation are calculated from
derivatives of the energy with respect to $\omega$ and $\lambda$, Eq.\
(\ref{LKDT}),
\FL
\begin{eqnarray}
\langle \hat{L}_{x} \rangle & = & -\hbar
(\frac{N_{y}}{\Omega_{+}}+\frac{N_{z}}{\Omega_{-}}) (\omega -
\frac{1}{2}(\frac{c}{b}+\frac{b}{c})\lambda ) -
\frac{\hbar}{2q}(\frac{N_{y}}{\Omega_{+}}-\frac{N_{z}}{\Omega_{-}})
\left[ 2(\omega_{y}^{2}+\omega_{z}^{2})\omega -
\frac{1}{2}((\frac{b}{c}+3\frac{c}{b})\omega_{z}^{2} +
(\frac{c}{b}+3\frac{b}{c})\omega_{y}^{2})\lambda \right] \nonumber \\
\langle \hat{C}_{x} \rangle & = & \hbar
(\frac{N_{y}}{\Omega_{+}}+\frac{N_{z}}{\Omega_{-}}) (\lambda -
\frac{1}{2}(\frac{c}{b}+\frac{b}{c})\omega ) +
\frac{\hbar}{2q}(\frac{N_{y}}{\Omega_{+}}-\frac{N_{z}}{\Omega_{-}})
\left[ (b^{2}+c^{2})(\frac{\omega_{y}^{2}}{c^{2}} +
\frac{\omega_{z}^{2}}{b^{2}})\lambda - \frac{1}{2}
((\frac{c}{b}+3\frac{b}{c})\omega_{y}^{2} +
(\frac{b}{c}+3\frac{c}{b})\omega_{z}^{2})\omega \right].
\end{eqnarray}
\narrowtext
To leading order in $\omega$ and $\lambda$, these exact quantum results
agree with the classical Riemann rotor formulas, as guaranteed by the theorem.
In particular, the quantum collective energy is given by
\begin{eqnarray}
E(\omega ,\lambda ) & = & \hbar (\Omega_{+}-\omega_{y})N_{y} + \hbar
(\Omega_{-
}-\omega_{z})N_{z} \nonumber \\
& & \mbox{} + (\omega \langle
\hat{L}_{x} \rangle - \lambda \langle \hat{C}_{x} \rangle ), \label{COLLE}
\end{eqnarray}
which equals $T(\omega ,\lambda )$ to quadratic order in $\omega$ and
$\lambda$.

\subsection{Applications}

Consider the case of $^{\rm 20}$Ne for which $N_{x}=N_{y}=14$, $N_{z}=22$.  If
$\omega =\lambda =0$, then the intrinsic energy is minimized, subject to a
constraint to constant volume, when the self-consistency condition is
satisfied,
Eq.\  (\ref{SELF}), or, equivalently,
$\omega_{i}=\omega_{0}(N_{x}N_{y}N_{z})^{1/3}/N_{i}$, where
$\omega_{0}^{3}=\omega_{x}\omega_{y}\omega_{z}$.  Fixing $\hbar
\omega_{0}=13.05$ MeV and $R=3.257$ fm implies the following initial data:
$\hbar \omega_{x}=\hbar \omega_{y}=15.175$ MeV, $\hbar \omega_{z}=9.656$
MeV, $a=b=3.093$ fm, $c=4.857$ fm.  If $\omega$ or $\lambda \neq 0$, then
the frequencies $\omega_{i}$ are chosen so as to minimize the intrinsic energy
in the rotating frame $\tilde{E}$, subject to the constraint of
constant volume, i.e.,
the product of the axes lengths is fixed, abc=46.465 fm$^{3}$.  In addition,
the
axes lengths must be self-consistent with the definition of the Kelvin
circulation,
viz., equality is achieved in Eqs. (\ref{AXES}).  The results of the
intrinsic energy
minimization are plotted in Figures (1-5).  The quantum angular momentum L
and the quantum Kelvin circulation C are given by the semiclassical expressions
$\langle L_{x} \rangle=\hbar \sqrt{L(L+1)}$ and $\langle C_{x} \rangle=\hbar
\sqrt{C(C+1)}$.  In Table I, self-consistent solutions that fit the $^{\rm
20}$Ne
experimental energy spectrum are tabulated.  Note that the classical Riemann
rotor kinetic energy $T(\omega ,\lambda )$, Eq.\  (\ref{RRKE}), gives good
agreement for small cranking velocities with the self-consistent quantum
collective energy $E(\omega ,\lambda )$, Eq.\  (\ref{COLLE}).

The most significant difference between the classical Riemann rotor formulas,
valid for small cranking velocities $\omega$ and $\lambda$, and the
 \mbox{self-consistent} quantum results is that the quantum rotational band is
cut
off.  This band termination is attained when the ellipsoid turns into an oblate
spheroid rotating about its symmetry axis, $a<b=c$.  At the cut off,
$\omega_{x}
> \omega_{y} = \omega_{z}$, $\Omega_{\pm}=\omega_{y}\pm |\omega -\lambda
|$ and the maximal angular momentum and circulation are attained
\begin{equation}
\langle \hat{L}_{x} \rangle = \langle \hat{C}_{x} \rangle = \frac{\omega -
\lambda}{|\omega -\lambda |}(N_{z}-N_{y}).
\end{equation}
In the $^{\rm 20}$Ne case, these maximal values are $N_{z}-N_{y}=8$.

For the low energy states of a rotational band in a heavy deformed nucleus, the
classical expressions are excellent approximations to the quantum cranking
formulas.  For example, in $^{\rm 166}$Er the total number of deformed
oscillator
quanta are $N_{x}=N_{y}=235$, $N_{z}=343$.  The band terminates when the
angular momentum $L = N_{z}-N_{y}=108$, and the classical results for a
Riemann rotor are excellent approximations when $L,C << 108$.  In Table II,
self-consistent results for this heavy deformed nucleus are presented; the
angular and vortex velocities are fit to the experimental energy and angular
momentum.  First, observe that the self-consistent collective energy $E(\omega
,\lambda )$ is well approximated by the classical value $T(\omega ,\lambda )$;
the error in the classical formula is less than $0.03\%$ up to angular momentum
$L=8$.  Second, the axes lengths are constant up to $L=8$.  Third, the rigidity
rises slowly with increasing angular momentum.  Finally, the ratio of the
Kelvin
circulation to the angular momentum increases from $0.42$ at $L=2$ to $0.51$
at $L=8$.

\section{CONCLUSION}

In this article, the quantum Riemann rotor model was formulated as a cranked
mean field theory.  If the mean field is approximated by the deformed
oscillator
potential, then self-consistent solutions correspond to classical Riemann
rotors
at small cranking velocities.  This semiclassical correspondence provides a
physical interpretation to the cranked angular $\omega$ and vortex $\lambda$
velocities of the quantum mean field theory.

In previous work in nuclear physics concerning the Riemann rotor model, an
algebraic model provided the framework for the quantum formulation of the
Riemann classical model.  What is the connection between the method of this
article and the prior algebraic work?

The algebraic framework provides a unifying perspective for the adiabatic
rotational model and the Riemann rotor model in both their classical and
quantum mechanical forms.  The two geometrical collective models are
associated with two subalgebras of the symplectic Sp(3,{\bf R}) Lie algebra,
known as ROT(3) and GCM(3),\cite{ROW85}
\begin{equation}
{\rm SO(3)} \subset {\rm ROT(3)} \subset {\rm GCM(3)} \subset {\rm Sp(3,{\bf
R})} .
\end{equation}
The rotational algebra ROT(3) is spanned by the one-body quadrupole operator
$Q^{(2)}$ plus the angular momentum algebra SO(3).  The general collective
motion Lie algebra GCM(3) is spanned by the full inertia tensor $Q$ plus the
general linear group Gl(3,{\bf R}).

The classical models corresponding to ROT(3) and GCM(3) are defined on the
phase spaces formed by coadjoint orbits in the duals of the Lie algebras of
ROT(3) and GCM(3).\cite{GUI80,MAR84,ROS79}  The Hamiltonian dynamical
systems on these coadjoint orbits are identical to the classical Euler rigid
rotor
model for ROT(3) and to the Riemann-Chandrasekhar-Lebovitz virial equations
of motion for GCM(3).\cite{ROS88}

The quantum models corresponding to ROT(3) and GCM(3) are created by
making a decomposition of the Fock space of antisymmetrized A-fermion states
into irreducible unitary representations of these two algebras.  These two
decompositions are achieved explicitly by making a change of variables to
collective and intrinsic coordinates.\cite{VIL70,ZIC71,ROW79,BUC79}  The
collective coordinates are defined on the orbits of the motion groups SO(3) and
Gl(3,{\bf R}) in {\bf R}$^{\rm 3A}$; the intrinsic coordinates are a smooth
transversal to the orbit manifolds.  For ROT(3), the irreducible representation
spaces correspond to the well-known adiabatic rotational
model.\cite{UI70,WEA73}  The GCM(3) decomposition into collective and
intrinsic coordinates is less familiar, because the end result is a poor
approximation to the physics of dynamical nuclear currents, i.e., the coupling
between the collective and intrinsic coordinates of GCM(3) is not
weak.\cite{ROS76,WEA76,ROS81}

A quantitative measure of the goodness of ROT(3) and GCM(3) symmetry is
obtained by their respective Casimir operators.  Within a single irreducible
representation, a Casimir operator is a multiple of the identity operator.
However, if ROT(3) or GCM(3) symmetry is a poor approximation and collective
nuclear states cannot by represented accurately as the product of collective
and
intrinsic wavefunctions, then their corresponding Casimir operators will not be
constant among the states of a rotational band.  There are two Casimirs of
ROT(3), $[Q^{(2)}\times Q^{(2)}]^{(0)}\propto \beta^{2}$ and $[Q^{(2)}\times
Q^{(2)}\times Q^{(2)}]^{(0)}\propto \beta^{3}\cos 3\gamma$, which measure the
deformation $\beta$ and triaxiality $\gamma$ of the inertia
ellipsoid.\cite{KUM72,MAC77,CLI88,ROS77}  The two ROT(3) Casimirs are
approximately constant if the nuclear shape is approximately constant, a good
first approximation to nuclear rotational bands.  Hence, ROT(3) symmetry is
useful for nuclear rotors, and the predictions of the adiabatic rotational
model
(Alaga rules for E2 transitions) are good first approximations to the
experimental
data.

The GCM(3) Casimir invariant $\hat{C}^{2} =\vec{C}\cdot \vec{C}$ is the squared
length of the Kelvin circulation vector.  The connection of this Casimir with
the
Kelvin circulation was not appreciated until recently;\cite{ROS88} $\vec{C}$
was
referred to as the vortex momentum by the discoverers of the GCM(3) Casimir
operator\cite{WEA76,GUL76}.  They established the following expression for it:
\begin{equation}
\hat{C}_{k} = \sum_{ij}\epsilon_{ijk}(\hat{Q}^{-1/2} \hat{N}
\hat{Q}^{1/2})_{ij}.
\label{KELVINOP}
\end{equation}
In the intrinsic rotating frame, $\hat{Q}$=diag($a^{2},b^{2},c^{2}$) is
diagonal
and the Gl(3,{\bf R}) generator is $\hat{N}_{ij}=\sum_{\alpha =1}^{A}x_{\alpha
i}p_{\alpha j}$.  Thus, the above definition specializes to Eq.\  (4)
in the intrinsic
frame coordinates.  In quantum mechanics, the net circulation is quantized to
nonnegative integral multiples of $\hbar$ and its squared length $\hat{C}^{2}$
to
$C(C+1)\hbar^{2}$.

This article's cranking calculations demonstrate that the GCM(3) Casimir is not
even roughly constant among the states of nuclear rotational bands.  Indeed,
for
a heavy deformed nucleus, it is the rigidity $r$ and axes lengths $a,b,c$ that
are
approximately constant; hence, the Kelvin circulation is approximately
proportional to the angular velocity, Eq.\  (\ref{Krw}).  Therefore, GCM(3)
symmetry is not found in real nuclei.  In the algebraic approach, this defect
is
remedied by extending the dynamical group to Sp(3,{\bf R}).\cite{ROW85}

Because of the complexity of the Kelvin circulation operator, Sp(3,{\bf R})
shell
model calculations to date have not attempted to determine the expectation of
this operator with respect to microscopic wavefunctions.  Heretofore, the only
microscopic information available about the Kelvin circulation operator is a
formal theorem that its eigenvalue spectrum within an infinite-dimensional
irreducible symplectic shell model space is identical to the angular momentum
spectrum of the associated $0\hbar \omega$ Elliott SU(3)
representation.\cite{ROS81,GEL73}  For an axially symmetric Sp(3,{\bf R}) and
SU(3) representation ($N_{x}=N_{y}$), the spectrum of C consists of the
nonnegative integers from $0$ to $N_{z}-N_{y}$.  This agrees with the cutoff of
this paper's mean field theory.

The Kelvin circulation operator is intractable in a shell model theory because
the
square root and the inverse of the inertia tensor are part of its definition,
Eq.\
(\ref{KELVINOP}).  The inertia tensor is only diagonal in the intrinsic frame,
a
property that shell model theory cannot exploit.  This paper's mean field
theory
of Riemann rotors takes advantage of the intrinsic frame to simplify the Kelvin
circulation operator by replacing the inertia tensor operator in the rotating
frame
by its c-number expectation, diag(a$^{2}$,b$^{2}$,c$^{2}$).  The inverse and
square root are then trivial.  This replacement is an approximation that
ignores
quantum shape fluctuations that are small for a deformed rotor compared to the
other terms in the Kelvin circulation operator.

\vspace{0.5cm}
This material is based upon work supported by the National Science Foundation
under Grant No.\ PHY-9212231.

\figure{The collective energy $E(\omega ,\lambda )$ of Eq.\  (\ref{COLLE}) is
plotted in MeV versus the angular $\hbar \omega$ and vortex $\hbar \lambda$
velocities in MeV. \label{FIG1}}

\figure{The quantum angular momentum L in units of $\hbar$ is plotted versus
the angular $\hbar \omega$ and vortex $\hbar \lambda$ velocities in MeV. Here
the quantum value $\sqrt{L(L+1)}$ is set equal to the semiclassical expectation
$\langle \hat{L}_{x} \rangle$. \label{FIG2}}

\figure{The quantum Kelvin circulation C in units of $\hbar$ is plotted versus
the
angular $\hbar \omega$ and vortex $\hbar \lambda$ velocities in MeV. Here the
quantum value $\sqrt{C(C+1)}$ is set equal to the semiclassical expectation
$\langle \hat{C}_{x} \rangle$. \label{FIG3}}

\figure{The expectation of the quadrupole operator $\langle Q_{20} \rangle =
\langle 2z^{2}-x^{2}-y^{2} \rangle$ in fm$^{2}$ is plotted versus the angular
$\hbar \omega$ and vortex $\hbar \lambda$ velocities in MeV. \label{FIG4}}

\figure{The expectation of the quadrupole operator $\langle Q_{22} \rangle =
\langle y^{2}-x^{2} \rangle$ in fm$^{2}$ is plotted versus the angular $\hbar
\omega$ and vortex $\hbar \lambda$ velocities in MeV. \label{FIG5}}

\bigskip
\widetext
\begin{table}
\squeezetable
\caption{Self-consistent calculation for the yrast rotational band in $^{\rm
20}$Ne}
\begin{tabular}{cccccccccc}
L&C&$\hbar \omega$ (MeV)&$\hbar \lambda$ (MeV)&E (keV)&T (keV)&$r$&a
(fm)&b (fm)&c (fm)\\
\tableline
0&\dec 0.0 &\dec 0.0 &\dec 0.0 &\dec 0.0 &\dec 0.0 & - &\dec 3.093 &\dec 3.093
&\dec 4.857 \\
2&\dec 0.921 &\dec 2.170 &\dec 1.539 &\dec 1.634 &\dec 1.640 &\dec .217
&\dec 3.090 &\dec 3.093 &\dec 4.857 \\
4&\dec 2.536 &\dec 3.143 &\dec 1.796 &\dec 4.247 &\dec 4.541 &\dec .371
&\dec 3.090 &\dec 3.103 &\dec 4.841 \\
6&\dec 4.120 &\dec 4.592 &\dec 2.383 &\dec 8.775 &\dec 10.527 &\dec .435
&\dec 3.084 &\dec 3.148 &\dec 4.780 \\
\end{tabular}
\label{Table1}
\end{table}
\bigskip

\bigskip
\widetext
\begin{table}
\squeezetable
\caption{Self-consistent calculation for the yrast rotational band in $^{\rm
166}$Er}
\begin{tabular}{cccccccccc}
L&C&$\hbar \omega$ (MeV)&$\hbar \lambda$ (MeV)&E (keV)&T (keV)&$r$&a
(fm)&b (fm)&c (fm)\\
\tableline
0&\dec 0.0 &\dec 0.0 &\dec 0.0 &\dec 0.0 &\dec 0.0 & - &\dec 5.927 &\dec 5.927
&\dec 8.651 \\
2&\dec 0.847 &\dec 0.112 &\dec 0.090 &\dec 80.57 &\dec 80.57 &\dec 0.136
&\dec 5.927 &\dec 5.927 &\dec 8.651 \\
4&\dec 1.858 &\dec 0.202 &\dec 0.162 &\dec 264.98 &\dec 264.99 &\dec 0.139
&\dec 5.927 &\dec 5.927 &\dec 8.651 \\
6&\dec 2.924 &\dec 0.289 &\dec 0.231 &\dec 545.44 &\dec 545.50 &\dec 0.143
&\dec 5.927 &\dec 5.927 &\dec 8.651 \\
8&\dec 4.043 &\dec 0.372 &\dec 0.296 &\dec 911.18 &\dec 911.38 &\dec 0.148
&\dec 5.927 &\dec 5.927 &\dec 8.651 \\
\end{tabular}
\label{Table2}
\end{table}


\begin{references}
\bibitem[1]{ROS88} G.\ Rosensteel, Ann.\ Phys.\ (N.Y.) {\bf 186}, 230 (1988).
\bibitem[2]{CHA69} S.\ Chandrasekhar, {\em Ellipsoidal Figures of Equilibrium}
(Yale University Press, New Haven, 1969) and references therein.
\bibitem[3]{LEB67} Norman R.\ Lebovitz, Ann.\ Rev.\ Astr.\ Astrophys.\ {\bf 5},
465 (1967).
\bibitem[4]{ING54} D.R.\ Inglis, Phys.\ Rev.\ {\bf 96}, 1059 (1954).
\bibitem[5]{RIN80} Peter Ring and Peter Schuck, {\em The Nuclear Many-Body
Problem} (Springer-Verlag, New York, 1980), Section 3.4.
\bibitem[6]{BOH75} Aage Bohr and Ben R. Mottelson, {\em Nuclear Structure}
(Benjamin, Reading, 1975) Vol.\ II.
\bibitem[7]{LUD60} G.\ L\"{u}ders, Z.\ Naturforsch.\ Teil A:{\bf 15}, 371
(1960).
\bibitem[8]{CUS68} R.Y.\ Cusson, Nucl.\ Phys.\ {\bf A114}, 289 (1968).
\bibitem[9]{GUE86} E.\ Moya de Guerra, Phys.\ Rep.\ {\bf 138}, 293 (1986).
\bibitem[10]{VAS87} M.G.\ Vassanji and D.J.\ Rowe, Nucl.\ Phys.\ {\bf A466},
227 (1987).
\bibitem[11]{ROS90a} G.\ Rosensteel,  Phys.\ Rev.\ C {\bf 41}, R811 (1990)
\bibitem[12]{GUE90} C.\ Garcia-Recio, T.W.\ Donnelly and E.\ Moya de Guerra,
Nucl.\ Phys.\ {\bf A}, 221 (1990).
\bibitem[13]{GUE88} D.\ Berdichevsky, P.\ Sarriguren, E.\ Moya de Guerra, M.\
Nishimura and D.W.L.\ Sprung, Phys.\ Rev.\ C {\bf 38}, 338 (1988).
\bibitem[14]{RIP86} G.\ Ripka and J-P.\ Blaizot, {\em Quantum Theory of Finite
Systems} (MIT Press, Cambridge, 1986), Section 3.2.
\bibitem[15]{RIP75} G.\ Ripka, J-P.\ Blaizot, and N.\ Kassis, Extended Seminar
on Nuclear Physics, Int.\ Center for Theor.\ Phys.\, Trieste, Paper
\mbox{IAEA-Smr-14/19}, 1975.
\bibitem[16]{STA78} A.P.\ Stamp, Z.\ Physik {\bf A284}, 305 (1978).
\bibitem[17]{FEY39}R.P.\ Feynman, Phys.\ Rev.\ {\bf 56}, 340 (1939).
\bibitem[18]{ROW85} D.J.\ Rowe, Rep.\ Prog.\ Phys.\ {\bf 48}, 1419 (1985) and
references therein.
\bibitem[19]{GUI80} V.\ Guillemin and S.\ Sternberg, Ann.\ Phys.\ (N.Y.) {\bf
127}, 220 (1980); {\em Symplectic techniques in physics} (Cambridge University
Press, Cambridge, 1984), Chapter II.
\bibitem[20]{MAR84} J.\ E.\ Marsden, T.\ Ratiu, and A.\ Weinstein, Trans.\
Amer.\ Math.\ Soc.\ {\bf 281}, 147 (1984).
\bibitem[21]{ROS79} G.\ Rosensteel and E.\ Ihrig, Ann.\ Phys.\ (N.Y.) {\bf
121},
113 (1979).
\bibitem[22]{VIL70} F.H.M.\ Villars and G.\ Cooper, Ann.\ Phys.\ (N.Y.) {\bf
56},
224 (1970).
\bibitem[23]{ZIC71} W.\ Zickendraht, J.\ Math.\ Phys.\ {\bf 12}, 1663 (1971).
\bibitem[24]{ROW79} D.J.\ Rowe and G.\ Rosensteel, Ann.\ Phys.\ (N.Y.) {\bf
126}, 198 (1980).
\bibitem[25]{BUC79} B.\ Buck, L.C.\ Biedenharn, and R.Y.\ Cusson, Nucl.\ Phys.
{\bf A317}, 205 (1979).
\bibitem[26]{UI70} H.\ Ui, Prog.\ Theor.\ Phys.\ {\bf 44}, 153 (1970).
\bibitem[27]{WEA73} L.\ Weaver, L.C.\ Biedenharn and R.Y.\ Cusson, Ann.\
Phys.\ (N.Y.) {\bf 77}, 250 (1973).
\bibitem[28]{ROS76} G.\ Rosensteel and D.J.\ Rowe, Ann.\ Phys.\ (N.Y.) {\bf
96},
1 (1976).
\bibitem[29]{WEA76} O.L.\ Weaver, R.Y.\ Cusson, and L.C.\ Biedenharn, Ann.\
Phys.\ (N.Y.) {\bf 102}, 493 (1976).
\bibitem[30]{ROS81} G.\ Rosensteel and D.J.\ Rowe, Phys.\ Rev.\ Lett.\ {\bf
46},
1119 (1981).
\bibitem[31]{KUM72} K.\ Kumar, Phys.\ Rev.\ Lett.\ {\bf 28}, 249 (1972).
\bibitem[32]{MAC77} Raymond S.\ Mackintosh, Rep.\ Prog. Phys.\ {\bf 40}, 731
(1977).
\bibitem[33]{CLI88} D. Cline, in {\em The Variety of Nuclear Shapes}, edited by
J.D.\ Garrett {\it et al.}  (World Scientific, Singapore, 1988), p.\ 1.
\bibitem[34]{ROS77}G.\ Rosensteel and D.J.\ Rowe, Ann.\ Phys.\ (N.Y.) {\bf
104}, 134 (1977).
\bibitem[35]{GUL76} P.\ Gulshani and D.J.\ Rowe, Can.\ J.\ Phys.\ {\bf 54}, 970
(1976).
\bibitem[36]{GEL73} S.\ Gelbart, Inv.\ Math.\ {\bf 19}, 49 (1973).
\end{references}
\end{document}